# Kinetic control for planar oxidation of MoS$_2$

Kate Reidy[1], Wouter Mortelmans[1], Seong Soon Jo[1], Aubrey Penn[2], Baoming Wang[1], Alexandre Foucher[1], Frances M. Ross[1], R. Jaramillo[1]

1. Department of Materials Science and Engineering, Massachusetts Institute of Technology, Cambridge, MA, USA

2. MIT.nano, Massachusetts Institute of Technology, Cambridge, MA, USA

**Abstract**

Layered transition metal dichalcogenide (TMD) semiconductors oxidize readily in a variety of conditions, and a thorough understanding of this oxide formation is required for the advancement of TMD-based microelectronics. Here, we combine scanning transmission electron microscopy (STEM) with spectroscopic ellipsometry (SE) to investigate oxide formation at the atomic scale of the most widely-studied TMD, MoS$_2$. We find that aggressive thermal oxidation results in α-phase plate-like crystalline MoO$_3$ with sharp interfaces, voids, and a textured alignment with the underlying MoS$_2$. Experiments with remote substrates and patterned MoS$_2$ prove that thermal oxidation proceeds via vapor-phase mass transport and redeposition - a challenge to forming thin, conformal planar oxide films. We accelerate the kinetics of oxidation relative to the kinetics of mass transport using a non-thermal oxygen plasma process, to form a smooth and conformal amorphous oxide. The resulting amorphous MoO$_3$ films can be grown several nanometers thick, and we calibrate the oxidation rate for varying plasma processing conditions. Our results illustrate how TMD semiconductor oxidation differs significantly from oxidation of legacy semiconductors, most notably silicon, and provide quantitative guidance for managing both the atomic scale structure and thin film morphology of oxides in the design and processing of MoS$_2$ semiconductor devices.

**Main Text**

Controlling the oxidation process is a crucial step in the fabrication of semiconductor devices. This is well-known from the evolution of legacy semiconductors such as silicon, where process control challenges explain the decades lag between MOSFET patents and commercialization. Control of the oxidation process is no less important for emerging semiconductors, such as layered and two-dimensional (2D) transition metal dichalcogenides (TMDs), which oxidize readily in a variety of conditions.[1–3] Oxidation chemistry is more complex for compound semiconductors, including layered and 2D materials, and a fundamental understanding of this oxide formation is crucial for the commercialization of TMD-based microelectronics.

Here we present an atomic-scale kinetic study of the oxide formation of MoS$_2$, the most widely studied TMD semiconductor. Layered MoS$_2$ exhibits a variety of useful physical and electrical properties, such as lubrication by virtue of inter-layer sliding, enhanced catalytic activity, and strong light-matter coupling.[4–9] MoS$_2$ is already in widespread use in heterogeneous catalysis and solid-state lubrication; it is also of interest in the microelectronics, chemical sensing, and photonics industries.[5,10–13] In many of these applications, uncontrolled oxidation is a limiting factor in device performance, while controlled oxidation can add functionality such as surface passivation, semiconductor-dielectric interfaces, and resistive switching.[14–17] Previous studies of MoS$_2$ show that it is resistant to oxidation at ambient and vacuum conditions, but that it will oxidize both in dry ($O_2$) and wet ($O_2 + H_2O$) thermal conditions.[17–22] Oxidation preferentially begins at defects such as chalcogen vacancies and grain boundaries, can be accelerated by illumination, and may be facilitated by organic absorbates.[23–28] $MoS_2$ oxidation has also been studied in environments typical of device processing, such as exposure to ozone, plasma, or wet etchants, and in these cases has been found to exhibit layer-thinning and self-limited amorphous oxide formation.[29–35] These reported TMD oxidation mechanisms depart strikingly from the more familiar examples of Si and Ge, and establishing kinetic control of these competing oxidation processes is crucial for future device integration of $MoS_2$.[36]

In this work, we use scanning transmission electron microscopy (STEM) and spectroscopic ellipsometry (SE) to describe the kinetics of oxidation of the basal plane of large $MoS_2$ crystals. We find that under thermal conditions, oxide formation is followed immediately by oxide desorption (*i.e.*, sublimation), resulting in a process of vapor-phase transport, re-deposition, and growth of α-phase crystalline $MoO_3$ (α-$MoO_3$) in textured alignment with the $MoS_2$ substrate. This process creates discontinuous films with large α-$MoO_3$ crystals and voids, and precludes the formation of a thin, conformal oxide that is often necessary in device fabrication. To oxidize the Mo atoms without subsequent mass transport in a topotactic growth mode – here termed 'oxidize-in-place' – it is necessary to use a non-thermal oxidation process, such as oxygen plasma processing. We find the oxide formed via this non-thermal route is amorphous and conformal to the substrate, in contrast to the thermal oxide. STEM confirms the amorphous structure of the plasma oxide, and electron energy loss spectroscopy (EELS) reveals that the band gap of amorphous $MoO_3$ is slightly reduced relative to the thermal, crystalline oxide. We discuss the implications of these distinct oxidation pathways in terms of atomic-scale structure and thin-film morphology to inform future device processing.

In **Fig. 1a** we show an example of a naturally occurring bulk $MoS_2$ crystal (2H structure, space group $P6_3/mmc$, #194) which has been thermally oxidized in a box furnace at 500 °C for 1 hour in 0.2 atm $O_2$ (**Methods**). Thermal oxidation results in the striking appearance of blocks of layered, crystalline α-$MoO_3$. Cross sectional STEM allows us to determine the total thickness of the oxide as approximately 230 nm (**Fig. 1b**). The oxide consists of multiple blocks, each 30-100 nm thick, which exhibit a plate-like morphology separated laterally by voids. These blocks have faceted sidewalls and are separated vertically by boundaries that appear darker in the STEM image (indicating regions of lower atomic number and/or lower density). This morphology is indicative of high degree of $MoO_x$ species mobility and a low density of oxide nucleation sites on the $MoS_2$ surface; we hypothesize both are related to the volatility of $MoO_x$. At higher magnification (**Fig. 1c-e**) we see the α-$MoO_3$ is textured (see also **Fig. S1b**), with orientation $MoO_3[010]||MoS_2[0001]$.

Composition measurements by energy-dispersive x-ray spectroscopy (EDS) indicate the presence of oxygen and a non-negligible quantity of sulfur in the oxide layer, as well as a sharp sulfide-oxide interface with no detectable chemical mixing (**Fig. 1f**). Probing further the oxidation state using EELS shows the 532 eV oxygen (O) K-edge and 382 eV molybdenum (Mo) M-edge (**Fig. 1g, Methods**). Comparison of this EELS data with reference spectra from literature shows peak positions and lineshapes indicative of $MoO_3$ (**Fig. S2**). This allows us to eliminate the possibility of other oxides, such as $MoO_2$, since they exhibit notable differences in Mo M-edge and O K-edge spectra.[37–39] A small shoulder in the oxygen K-edge indicates the presence of oxygen vacancies, consistent with a defective $MoO_3$ structure with a low level of sulfur incorporation.[40,41]

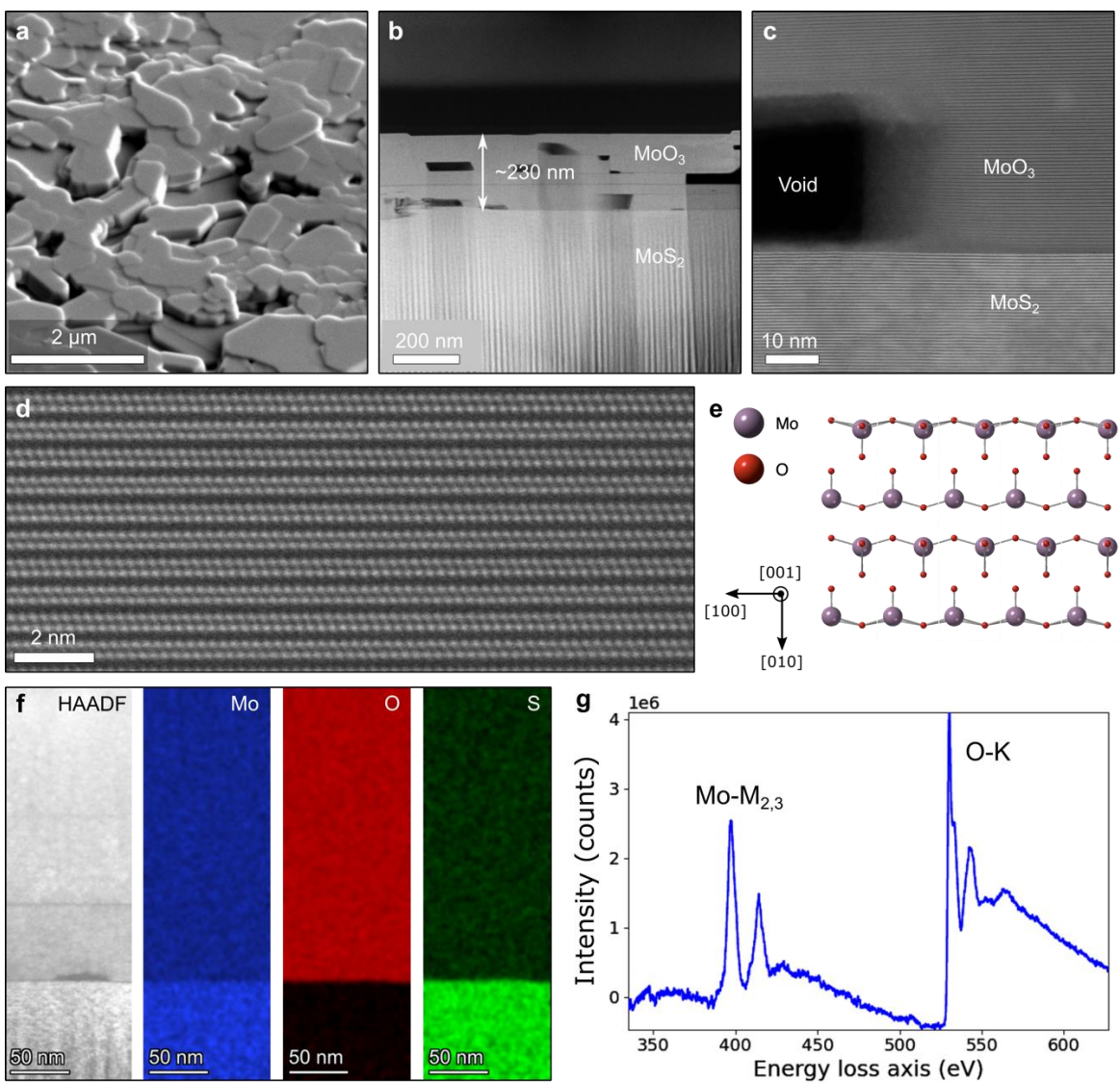

**Figure 1: Characterization of the α-MoO₃ crystalline thermal oxide.** (a) Scanning electron microscopy (SEM) image of freshly exfoliated bulk MoS$_2$ oxidized at 500 °C for 1 hour in 0.2 atm O$_2$. (b) Cross sectional STEM image and (c), atomic resolution STEM image of the same sample. The darker region beneath each void and the vertical wrinkles in the substrate are curtaining artifacts caused by focused ion beam (FIB) preparation. (d) Atomic resolution STEM HAADF image of MoO$_3$ lattice (Mo atoms visible) and (e) corresponding atomic model looking down the MoO$_3$ [001] zone axis. We use conventional indexing for α-MoO$_3$, with the long axis along $\hat{b}$. (f) EDS elemental mapping of Mo, O and S distribution in the sample. (g) EEL spectrum of the oxide region showing Mo-M$_{2,3}$ and O-K edges. Area of EEL spectrum collection and further details of Mo-M and O-K edges shown in **Fig. S2**.

To determine the mechanism of α-MoO$_3$ crystal growth, we performed experiments with remote and patterned substrates, as shown in **Fig. 2**, where: (1) freshly-exfoliated MoS$_2$ is sandwiched between pre-cleaned Si/SiO$_2$ wafers, with the top Si wafer placed gently on the MoS$_2$ flake resulting in micron-scale gaps (**Figs. 2a-b**), and (2) mechanically exfoliated graphite is transferred onto the MoS$_2$ and used as a patterned mask (**Figs. 2c-d**). In both cases, thermal oxidation resulted in MoO$_3$ crystallites grown on the remote and patterned substrates. This indicates that thermal oxidation occurs via vapor-phase mass transport and redeposition. We propose that oxidation of the MoS$_2$ surface is followed immediately by sublimation of a volatile molecular species MoO$_x$ ($x$ unknown). The local vapor pressure of MoO$_x$ is then a source for vapor-phase deposition and growth of crystalline α-MoO$_3$, on the original substrate and other nearby surfaces. This is in sharp contrast to the process by which SiO$_2$ forms by oxidation of Si, with the formation of an amorphous oxide and subsequent diffusion of oxygen through the oxide to the Si reaction front (the Deal-Grove model). This vapor phase mass transport mechanism explains why the thermal oxide layers reported here resemble films formed by the thermal evaporation of MoO$_3$ on various layered materials via van der Waals epitaxy.[42–44] Further evidence of the vapor phase mass transport and redeposition mechanism is shown in **Fig. 2b**, where the bright white areas are crystals that nucleated from the vapor phase with a growth direction perpendicular to the substrate. This morphology is similar to that of MoO$_3$ grown on SiO$_2$ substrates via MoO$_3$ vapor phase evaporation.[43,45]

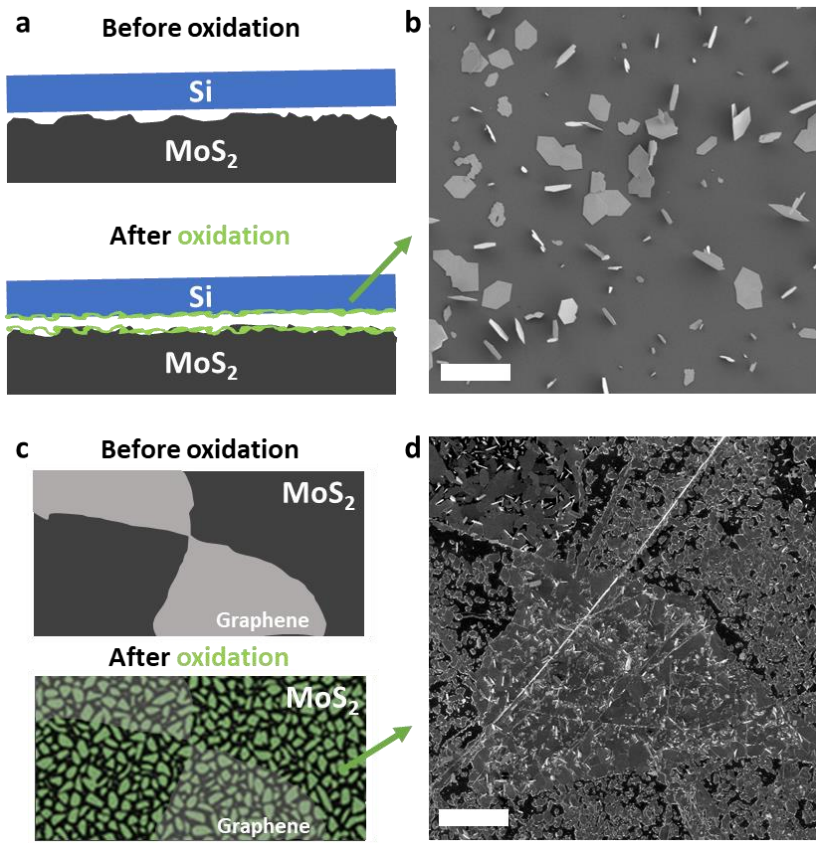

**Figure 2: Determining the mechanism of α-MoO$_3$ crystal growth.** Schematics illustrate experimental design; SEM images were recorded after oxidation. **(a-b)** MoS$_2$ crystal is sandwiched between pre-cleaned Si/SiO$_2$ wafers with a micron-scale gap. **(c-d)** Mechanically-exfoliated graphite flakes are transferred onto the MoS$_2$. Both samples are annealed at 500 °C and 0.2 atm O$_2$ for 1hr. Scale bars = 20 $\mu$m.

Most uses of MoS$_2$/MoO$_3$ heterojunctions in microelectronics and optoelectronics will require smooth interfaces, to enable planar processing and uniform devices. We therefore aim to tailor the MoO$_3$ nucleation process and resulting oxide morphology by controlling the oxidation environment. To minimize MoO$_x$ sublimation, we first carried out rapid thermal annealing (RTA) at a fast rate of heating (20 °C/sec), reducing material loss during the ramp. Even at lower temperature (450 °C) and shorter oxidation times (1 min, 5 min, and 10 min) we still observe morphology that suggests the mechanism of sublimation and re-deposition (**Figs. 3a-c**). The initial shape of the oxide crystals is rectangular, which is consistent with the morphology of ultrathin textured MoO$_3$ deposited from the vapor phase (**Fig. 3a**).[14,42,43] This is followed by anisotropic growth along preferential growth directions [100] or [001], the formation of elongated hexagonal shapes (**Fig. 3b**), and lastly by coalescence (**Fig. 3c**).[45,46]

Given the volatility of MoO$_x$, we hypothesize that increasing the oxygen partial pressure in the chamber could decrease the sublimation rate, and thus help in forming thin and conformal MoO$_3$ layers. The results of **Figs. 1-2** were obtained by oxidation in oxygen pressure $P(O_2) = 0.2$ atm, equivalent to 1 atm of air. We therefore oxidize MoS$_2$ at higher oxygen pressure, $P(O_2) = 0.4$ atm in **Figs. 3d-f**. Higher oxygen pressure does decrease MoO$_x$ volatilization, enabling in-plane oxide growth with higher surface coverage that increases with growth time. However, some volatilization still occurs, and micrometer-scale gaps remain in the film after 30 minutes of oxidation, as shown in **Fig. 3f**.

It is clear from the above that thermal processing will be challenged to produce a smooth and conformal oxide, due to high volatility of MoO$_x$ (**Fig. 3g, left**). Therefore, we must instead move towards an oxidize-in-place mechanism by accelerating the kinetics of oxidation relative to the kinetics of mass transport (**Fig. 3g, middle, right**). Since the rate of MoO$_x$ volatilization seems to be much higher than that of oxidize-in-place mechanisms at temperatures above 400 °C, yet at 350 °C and below thermal oxide growth is prohibitively slow (**Fig. S1, S3**), we infer that creating a smooth conformal oxide will require non-thermal processing.

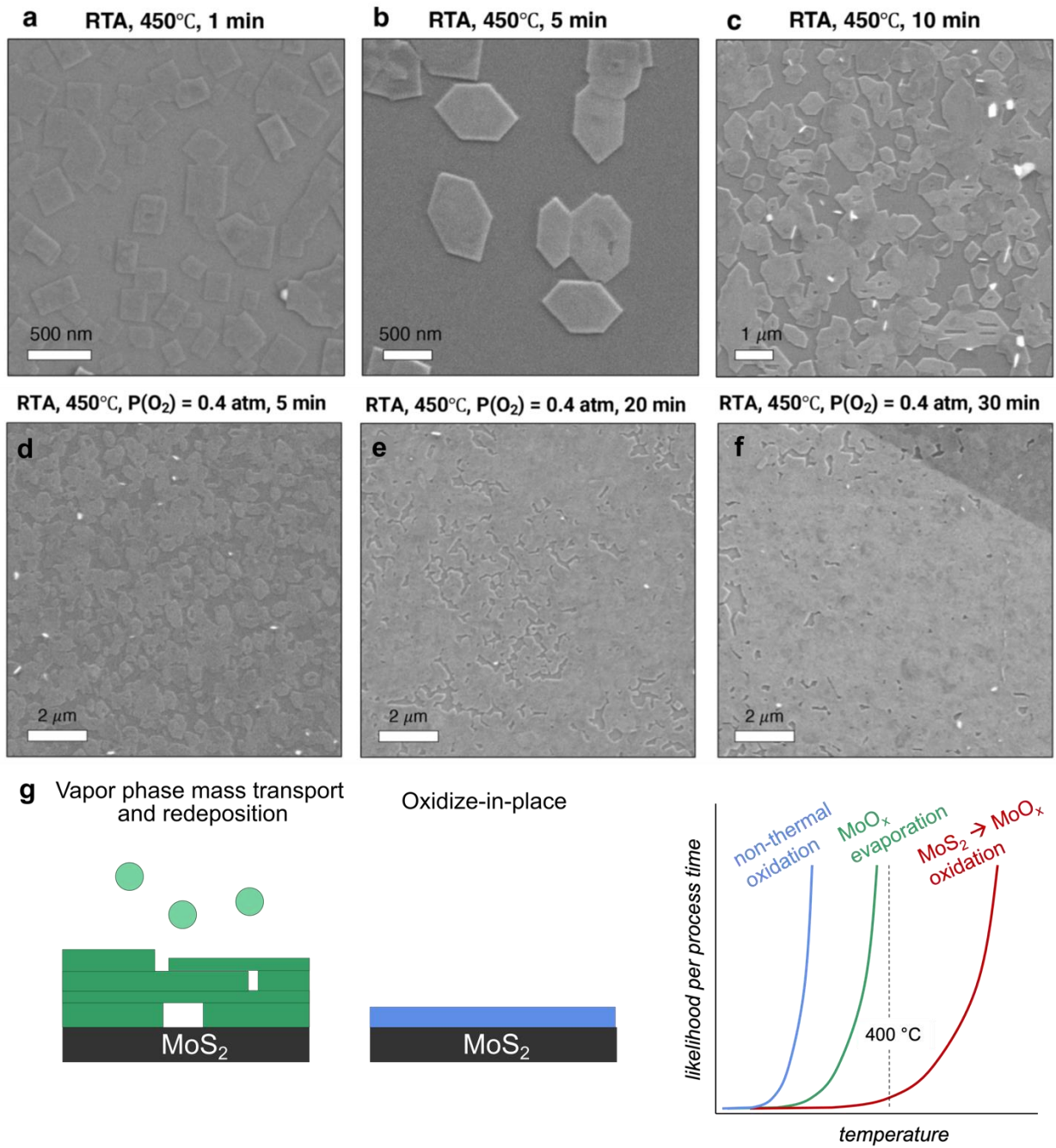

**Figure 3: Controlling MoS₂ oxidation and MoO₃ film morphology** (a-c) SEM images of freshly-exfoliated MoS$_2$ crystals oxidized at 450 °C and 0.2 atm O$_2$ for **(a)** 1 min, **(b)** 5 min and **(c)** 10 min. (d-f) SEM images of freshly-exfoliated MoS$_2$ crystals oxidized at 450 °C and 0.4 atm O$_2$ for (a) 5 min, (b) 20 min and (c) 30 min. **(g)** Schematic of the oxide film structures that result from vapor-phase mass transport *vs*. an oxidize-in-place process, and schematic of the kinetics of the key mechanisms: thermal oxidation of MoS$_2$ to MoO$_x$, evaporation of MoO$_x$, and non-thermal oxidation.

Oxygen plasma processing is a promising candidate to produce the desired oxidize-in-place kinetics for conformal MoO$_3$ growth on the few-nanometer scale. UV-ozone oxidation of MoS$_2$ can also produce conformal amorphous MoO$_3$ layers, although in this case the oxide thickness is self-limited to one monolayer.[32] **Fig. 4a-b** shows a cross-section of an MoS$_2$/MoO$_3$ heterostructure produced via radiofrequency (RF) oxygen plasma processing (**Methods**). Cross-sectional STEM (**Fig. 4a**) confirms that this oxide is thin (~4-7 nm, see also **Fig. S4**), and conformal with the underlying MoS$_2$. Furthermore, unlike the crystalline oxide formed via thermal oxidation, the plasma-processed oxide is amorphous. **Figs. 4c, d** show STEM nanodiffraction patterns measured within the MoS$_2$ crystal and the amorphous MoO$_3$ layer. The oxide exhibits no diffraction spots, indicating an amorphous structure. Therefore, control of oxide morphology *and phase* can be achieved by choice of oxidation method: thermal oxidation produces crystalline α-MoO$_3$ with uneven and difficult-to-control morphology, whereas RF oxygen plasma produces thin and conformal amorphous MoO$_3$. The crystalline and amorphous phases are determined to have slightly different band gap: 4.0 ± 0.1 eV for α-MoO$_3$, compared to 3.6 ± 0.1 eV for amorphous MoO$_3$, estimated from linear fits to the conduction band onset in EELS spectra (**Fig. 4e, Methods**).[47,48] Both results are consistent with MoO$_3$ band gap values found in literature. The measured properties of MoO$_3$, including the apparent band gap, vary strongly with oxygen vacancy concentration; our results show that crystallinity is another factor that controls the band gap.[17,44,49,50]

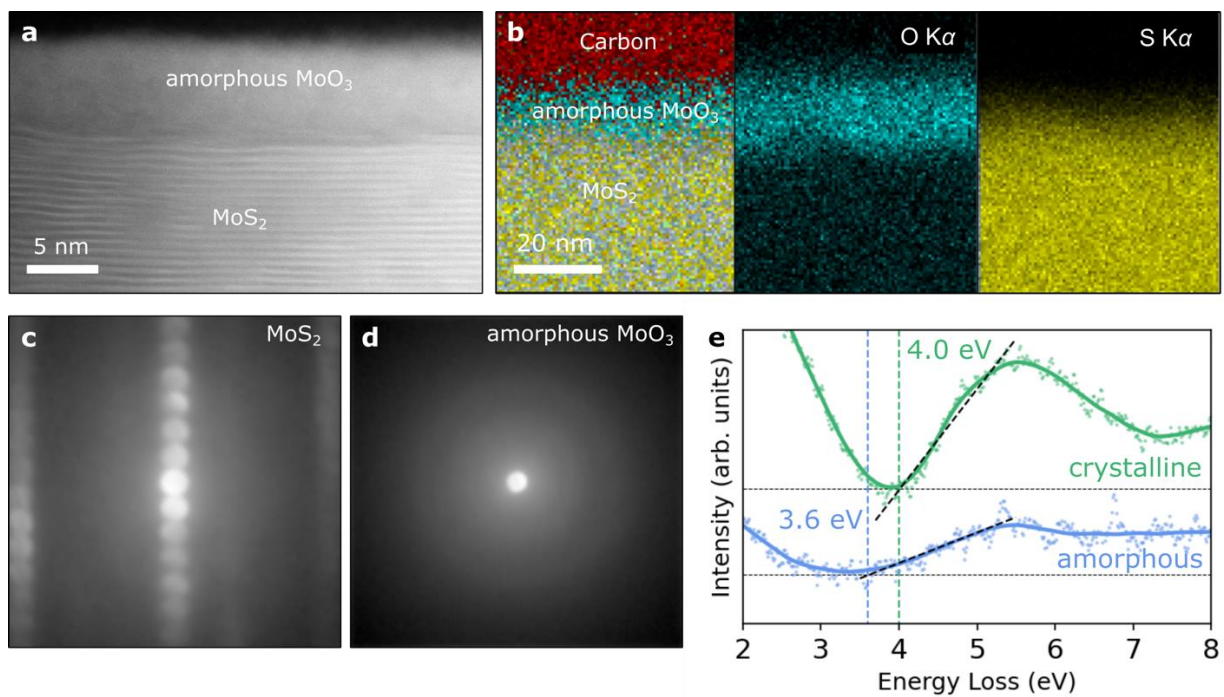

**Figure 4: Morphology, structure, and composition of plasma-processed amorphous MoO$_3$**. **(a)** Cross sectional STEM image demonstrating a thin, conformal oxide. **(b)** STEM-EDS element mapping of combined signal, O K$\alpha$, and S K$\alpha$ distribution within the sample. **(c, d)** Convergent beam electron diffraction (CBED) pattern from a STEM nanodffraction dataset averaged over the MoS$_2$ crystal and the amorphous MoO$_3$ film, respectively. **(e)** EELS data measured on thermal α-MoO$_3$ and plasma amorphous MoO$_3$. Points show raw data, while lines indicate data denoised data via Lowess smoothing. The band gap is determined by the intersection of the linear fit to conduction band onset with the quasi-elastic background (further details on EELS

analysis in **Methods**). The crystalline $MoO_3$ spectrum is offset vertically from the amorphous $MoO_3$ spectrum for visualization.

We see in **Fig. 4** that plasma processing can produce films of amorphous $MoO_3$ that are much thicker than a single layer of Mo-O bonds or a single unit cell. We use spectroscopic ellipsometry (SE) to measure how amorphous $MoO_3$ layer thickness varies with oxidation time and plasma conditions. In **Fig. 5a** we present the variation in amorphous $MoO_3$ thickness for processing times up to 30 min, plasma power varying between 100 – 250 W, and $O_2$ flow between 1 – 3 sccm. The raw SE data and fits are presented in SI (**Fig. S5**). The oxidation process appears to be self-limiting, judging from the near-constant oxide thickness for processing times of 20 min and beyond. In the most aggressive conditions (3 sccm $O_2$, plasma power 250 W, chamber pressure 1.3 torr), the oxide grows to 4.8 ± 0.2 nm after 30 min. More gentle oxidation conditions (1 sccm, 100 W, 0.7 torr) produce a thinner, self-limited oxide thickness of 3.2 ± 0.1 nm.

The oxidation rate is initially very fast and is not fully captured by our sequence of processing times. We attribute this in part to hardware limitations. In the processing tool used (an asher), the plasma is ignited close to the $MoS_2$ surface, and the sample temperature is uncontrolled. We hypothesize that rapid changes in $O_2$ pressure, plasma conditions, and sample temperature at the start of processing result in rapid oxidation. A different tool, with the ability to controllably approach the sample towards a remote plasma, may yield finer control of TMD oxidation.

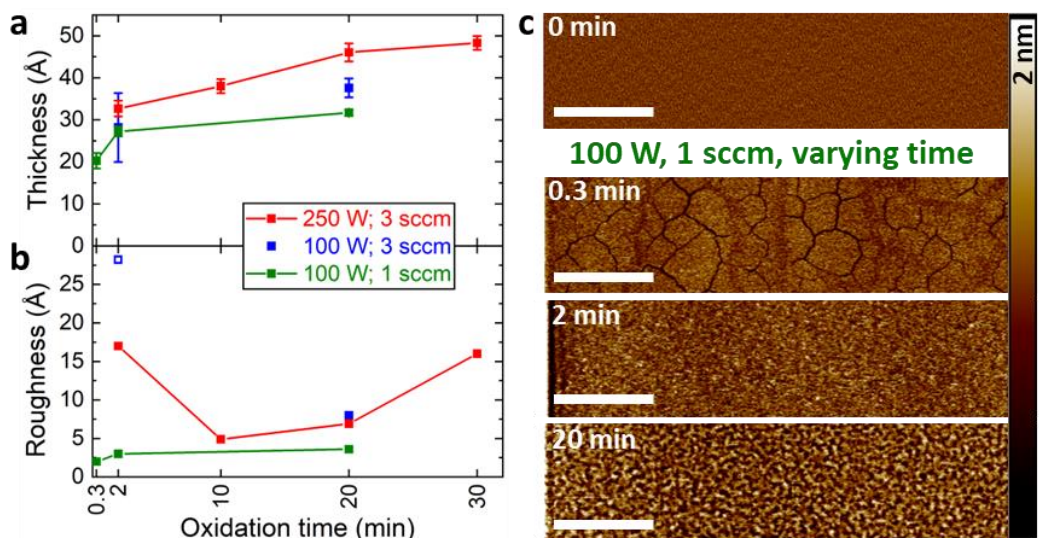

**Figure 5: Characterizing the $MoS_2$ plasma oxidation process.** **(a)** Oxide thickness and **(b)** surface roughness for varying processing time and plasma conditions. The oxide thickness is obtained from SE data modelling. Surface roughness is included in the model, using the effective medium approximation, and is set to the roughness measured by AFM (except for the open symbol). **(c)** AFM micrographs of the sample surface for the time series of gentle oxidation conditions (1 sccm $O_2$, 100 W); scale bars are 1 µm.

The most aggressive plasma conditions and fast initial oxidation lead to substantial surface roughening, which we measure using atomic force microscopy (AFM), and present in **Figs. 5b-c**. The $MoS_2$ surface is at first atomically smooth (initial roughness ≈ 0.1 nm). Plasma processing initially produces a rough oxide surface (roughness > 1.5 nm). As processing continues, the roughness decreases before gradually increasing. The final roughness, at the self-limited thickness, may explain the few nanometer thickness variation observed in cross sectional STEM (**Fig. S4**). For more gentle plasma conditions, the surface remains atomically smooth as oxidation begins, and shows only a gentle increase of roughness to 0.36 nm after 20 min (oxide thickness $3.2 \pm 0.1$ nm) (**Fig. 5c**). Upon initial conversion from $MoS_2$ to $MoO_3$, nanoscopic cracks appear in the oxide. This can be explained by the higher density of Mo atoms in the oxide compared to the sulfide, combined with limited mass transport on the nominally unheated sample surface. Thermal expansion due to localized heating from the plasma ignition can also have an effect on the surface morphology. Upon further oxidation, these cracks become filled in.

In summary, we have identified that thermal oxidation of $MoS_2$ proceeds via oxide formation, sublimation, vapor-phase mass transport, and redeposition, resulting in textured growth of crystalline α-$MoO_3$. These complicated kinetics follow from the fact that $MoO_x$ is highly volatile (unlike $SiO_2$). Having understood these kinetics, we propose and demonstrate that non-thermal oxidation by oxygen plasma processing can produce a smooth and conformal amorphous planar oxide, via a kinetically-limited oxidize-in-place process. Plasma processing can form amorphous $MoO_3$ layers up to several nanometers thick, with a band gap slightly reduced from crystalline α-$MoO_3$. We report the growth rate and roughness for varying plasma processing time and conditions. Further work on oxidation kinetics may include chemical analysis of the vapor-phase oxide species. We also look forward to studies of how processing conditions affect the properties of amorphous $MoO_3$ for particular applications, including hole transport, passivation, and resistive switching.

The guidance presented here for managing oxide atomic structure and thin-film morphology can aid in the design and processing of devices with $MoS_2$/$MoO_3$ heterojunctions. For example, crystalline islands of $MoO_3$ (thermally processed) may be preferred in applications such as catalysis, where the surface area and crystal facet are important factors - thermal oxidation of $MoS_2$ has been demonstrated here as an efficient way to create faceted $MoO_3$ islands on an arbitrary mask layer. Concurrently, amorphous $MoO_3$ thin films (non-thermal processing) are preferred for nonvolatile resistive switching. For some applications, plasma processing may be followed by annealing to convert the $MoO_3$ from its amorphous to crystalline forms while maintaining conformal thin-film morphology. By tailoring the growth kinetics, both crystalline and amorphous phases as well as islanded and thin-film morphologies can be accessed via $MoS_2$ oxidation.

**Experimental details**

Sample sourcing and preparation

We study naturally occurring $MoS_2$ crystals (2H structure, space group: $P6_3/mmc$, #194) received from the Smithsonian Institution (catalog number NMNH B3306). We prepare free-standing $MoS_2$ sample of ≤ 1 cm x 1 cm by cleaving the bulk counterpart using single-edge razor blade. We use adhesive tape (3M, Scotch) to create newly exposed and oxide-free surfaces by mechanical cleaving.

Thermal oxidation

We oxidize $MoS_2$ between 350 to 500 °C in a laboratory box furnace in laboratory ambient conditions, with relative humidity of ≈50%. The heating rate is 10 °C/min. The processing temperature is held for varying durations (1, 5, 30 and 60 min). After the high-temperature stage is complete, the furnace is switched off allowed to cool uncontrolled to room temperature.

Rapid thermal annealing (RTA)

We use an RTA system (AS-One, ANNEALSYS), with has a heating and cooling rate of 20 °C/sec and 1 °C/sec, respectively. We load the cleaved $MoS_2$ sample directly onto the susceptor, followed by pumping the chamber down to < 5 torr. The chamber is then brought to atmospheric pressure with a gas mixture of $N_2$ and $O_2$. We control $P(O_2)$ by varying the ratio of gas flows. For instance, we achieve $P(O_2) = 0.4$ with $N_2$ and $O_2$ flows of 600 and 400 sccm, respectively. The RTA chamber is heated to 450 °C for oxidation at varying durations (1, 5, 10, 20 and 30 min). After the high-temperature stage is complete, the chamber is cooled to 100 °C, at which point the chamber can be opened and the sample unloaded.

Oxygen plasma processing

We use an oxygen plasma system (AutoGlow) with an unheated sample (nominally at room temperature). The $MoS_2$ crystal is placed in the chamber, and the chamber is pumped down to < 0.5 Torr before introducing $O_2$ gas and igniting the plasma.

Spectroscopic ellipsometry (SE)

We perform SE measurements on the (001) basal planes (perpendicular to the optic axis) in the photon energy range 1.5 to 4.1 eV (800 to 300 nm). The ellipsometer (UV-NIR Vase, from J. A. Woollam) uses an automatic rotating analyzer and an auto-retarder. We perform all measurements on mirror-smooth surfaces. We carry out all measurements at an angle-of-incidence of 65 and 70°. For clarity, only the 70º angles are presented.

We use SE analysis software to perform model-based analysis of the SE data. We use an optical model consisting of four layers – air, surface roughness, thin oxide layer, and semi-infinite bulk crystal. We assume the dielectric properties of the oxide and underlying $MoS_2$ crystal do not change over time. We model oxide surface roughness using the effective medium approximation, with the roughness being fixed by AFM measurements on the same samples. We use $(n,\kappa)$ data for $MoS_2$ from SE measurements on our pristine bulk surfaces. We use a Cauchy model for the optical properties of $MoO_3$. We perform best-fit regression to the data for varying oxide thickness and Cauchy model parameters. The raw data and fittings are presented in **Fig. S5**.

Scanning electron microscopy (SEM)

We used a high- resolution SEM (Zeiss Merlin) for SEM imaging. The SEM plan-view images are collected mostly in secondary electron imaging mode, except for graphene/$MoS_2$ samples, which were examined using an in-lens detector for higher image resolution.

Scanning transmission electron microscopy (STEM)

We performed STEM imaging and convergent beam electron diffraction on a probe-corrected Thermo Fisher Scientific Themis Z G3 60–300 kV S/TEM operated at 200 kV with a beam current of 30-40 pA and 25 mrad convergence angle. We acquired EDS spectra for 1 hour with drift correction at 1,000 – 3,000 counts/second. We collected STEM nanodiffraction (4D STEM) data in microprobe STEM mode using an electron microscope pixel array detector (EMPAD), with a convergence angle of 1 mrad and a dwell time of 1 msec. We prepared samples for cross-sectional

TEM characterization via standard liftout techniques, using a focused ion beam (FEI Helios 600 and FEI Helios 660 Dual-Beam FIB/SEM) system with a carbon protection layer. Samples were further thinned and polished using a Fischione Nanomill 1040. We acquired EDS data for the amorphous oxide sample on a JEOL ARM 200-CF operated at 80kV with an EDAX Octane 100mm$^2$ detector.

We acquired EEL spectra at 200 kV on a Thermo Fisher Scientific Themis Z G3 60–300 kV S/TEM. For the high-loss EELS measurements, the energy resolution (measured as the full width at half maximum (FWHM) of the zero loss peak (ZLP)) was 0.7 eV with an energy dispersion of 0.15 eV/ch (overview spectra), and 0.05 eV/ch (fine structure). The 0.15 eV dispersion allowed us to record Mo-M$_{2,3}$ and O-K edges in the same spectra. Each crystal was set slightly off zone axis to avoid channeling effects. Some aliasing effects were also observed, which we could avoid by applying defocus. Dwell time of the EELS acquisition was set short enough to avoid Mo reduction under the electron beam, as previously reported for Mo oxides.[41] .We carried out the low-loss EELS band gap measurements with a monochromated beam at 0.2 eV FWHM, 0.015 eV/ch dispersion, 50 – 120 pA current, 40 mrad convergence angle, and exposure time of 0.02 – 0.05 s.

STEM imaging and STEM-EDS shown in **Fig. 4b** were performed on a JEOL ARM 200-CF operated at 80 kV, and using an EDAX Octane 100 mm$^2$ detector. This sample was prepared following standard liftout techniques, using an FEI Helios 660 Dual-Beam FIB/SEM, and further thinned and polished using a Fischione Nanomill 1040.

EELS data analysis

For the high-loss regime: All spectra were first x-ray removed and aligned by the ZLP using the Gatan Microscopy Suite. We performed background subtraction in the pre-edge energy window for each spectrum. We then removed the multiple scattering by Fourier-ratio deconvolution with the aligned low-loss spectrum obtained for the same region of the sample. A relative measure of the specimen thickness in units of the local inelastic mean free path (τ/λ) varied from 0.20 – 0.27 across the sample, indicating a relatively thin sample.

For the low-loss (conduction band onset) regime: All spectra were first x-ray removed and aligned by the ZLP using the Gatan Microscopy Suite. We denoised the data from the area of interest using a LOWESS (Locally Weighted Scatterplot Smoothing) algorithm, implemented in hyperspy.[51] The smoothing parameters used were: 0.15 for the crystalline data, and 0.197 for the amorphous data. Both raw data and smoothed data are plotted in **Fig. 4**. We then obtained a line of best fit for the conduction band onset using a basic polyfit function. The slope and intercept of this line can vary depending on the data segment chosen for fitting, and we estimated the error in bandgap as ± 0.1 eV to account for this.

X-ray Photoelectron Spectroscopy (XPS)

We characterized chemical states using XPS (Nexsa G2 Surface Analysis System, ThermoFisher Scientific) with a monochromatic aluminum Kα X-ray source (spot size ~ 100 µm), and pass energy of 50 eV for high-resolution scans. Before collecting XPS spectra, we used a monatomic Ar ion beam with 1000 eV for 30 s to remove surface adsorbates such as hydrocarbons.

Mechanically exfoliated graphite and transfer

We exfoliated graphite onto a pre-cleaned Si/SiO$_2$ substrate using scotch tape and adhesive tape (Nitto Denko America Inc. 3195MS). The exfoliated graphite flake is then transferred onto MoS$_2$ using a cellulose acetate butyrate (CAB)-mediated transfer method.[52,53]

**Acknowledgements**

This work was funded in part by Semiconductor Research Corporation (SRC). This work was supported by the Office of Naval Research (ONR) MURI through Grant No. N00014-17-1-2661. This work was carried out with the use of facilities and instrumentation supported by NSF through the Massachusetts Institute of Technology Materials Research Science and Engineering Center DMR - 1419807. This work was carried out in part through the use of MIT.nano's facilities. This work was performed in part at the Harvard University Center for Nanoscale Systems (CNS). K.R. acknowledges funding and support from a MIT MathWorks Engineering Fellowship and ExxonMobil Research and Engineering Company through the MIT Energy Initiative. We thank Prof. Deji Akinwande (University of Texas at Austin), and Drs. Sudarat Lee, Kevin O'Brien, and Chelsey Dorow (Intel Corporation) for useful discussions. We thank Dr. Yifei Li (MIT), and Dr. Austin Akey and Dr. Jules Gardener (Harvard University) for technical assistance.

**Conflict of Interest**

The authors declare no conflict of interest.

**Author Contributions**

R.J. and F.M.R. conceived and directed the project. K.R. performed STEM imaging, EDS mapping, and EELS data analysis. W.M. performed plasma oxidations, spectroscopic ellipsometry and atomic force microscopy. S.J. performed SEM imaging, thermal oxidations including RTA studies, EDS on plasma oxide, XPS, and Raman. K.R. and S.J. exfoliated and transferred graphene. A.P. and K.R. performed EELS mapping, atomic resolution STEM, and 4D STEM. B.W. and A.F. performed cross-sectional FIB cuts. All authors have given approval to the final version of the manuscript.

**References**

1. Ross, S. & Sussman, A. Surface oxidation of molybdenum disulfide. *J. Phys. Chem.* **59**, 889–892 (1955).

2. Li, Q., Zhou, Q., Shi, L., Chen, Q. & Wang, J. Recent advances in oxidation and degradation mechanisms of ultrathin 2D materials under ambient conditions and their passivation strategies. *J. Mater. Chem. A* **7**, 4291–4312 (2019).

3. Kc, S., Longo, R. C., Wallace, R. M. & Cho, K. Surface oxidation energetics and kinetics on MoS$_2$ monolayer. *J. Appl. Phys.* **117**, 135301 (2015).

4. Mak, K. F., Lee, C., Hone, J., Shan, J. & Heinz, T. F. Atomically thin MoS$_2$: A new direct-gap semiconductor. *Phys. Rev. Lett.* **105**, 136805 (2010).

5. Butler, S. Z. *et al.* Progress, Challenges, and Opportunities in Two-Dimensional Materials Beyond Graphene. *ACS Nano* **7**, 2898–2926 (2013).

6. Kunstmann, J. *et al.* Momentum-space indirect interlayer excitons in transition-metal dichalcogenide van der Waals heterostructures. *Nat. Phys.* **14**, 801–805 (2018).

7. Roch, J. G. *et al.* Spin-polarized electrons in monolayer MoS$_2$. *Nat. Nanotechnol.* **14**, 432–436 (2019).

8. Shen, P. C. *et al.* Healing of donor defect states in monolayer molybdenum disulfide using oxygen-incorporated chemical vapour deposition. *Nat. Electron.* **5**, 28–36 (2022).

9. Splendiani, A. *et al.* Emerging photoluminescence in monolayer MoS2. *Nano Lett.* **10**, 1271–1275 (2010).

10. Akinwande, D., Petrone, N. & Hone, J. Two-dimensional flexible nanoelectronics. *Nat. Commun.* **5**, 5678 (2014).

11. Liu, X. & Hersam, M. C. 2D materials for quantum information science. *Nat. Rev. Mater.* **4**, 669–684 (2019).

12. Choi, G. J. *et al.* Polarized Light-Emitting Diodes Based on Patterned MoS$_2$ Nanosheet Hole Transport Layer. *Adv. Mater.* **29**, 1702598 (2017).

13. Choi, W. *et al.* Recent development of two-dimensional transition metal dichalcogenides and their applications. *Mater. Today* **20**, 116–130 (2017).

14. Cai, L. *et al.* Rapid flame synthesis of atomically thin MoO$_3$ down to monolayer thickness for effective hole doping of WSe$_2$. *Nano Lett.* **17**, 3854–3861 (2017).

15. Kidambi, P. R., Weijtens, C., Robertson, J., Hofmann, S. & Meyer, J. Multifunctional oxides for integrated manufacturing of efficient graphene electrodes for organic electronics. *Appl. Phys. Lett.* **106**, 063304 (2015).

16. Chuang, S. *et al.* MoS$_2$ p-type transistors and diodes enabled by high work function MoO$_x$ contacts. *Nano Lett.* **14**, 1337–1342 (2014).

17. Bessonov, A. A. *et al.* Layered memristive and memcapacitive switches for printable electronics. *Nat. Mater. 2014 142* **14**, 199–204 (2014).

18. Soon Jo, S. *et al.* Growth Kinetics and Atomistic Mechanisms of Native Oxidation of ZrS$_x$Se$_{2-x}$ and MoS$_2$ Crystals. *Nano Lett.* **120**, 8592–8599 (2020).

19. Wu, J. *et al.* Layer Thinning and Etching of Mechanically Exfoliated MoS$_2$ Nanosheets by Thermal Annealing in Air. *Small* **9**, 3314–3319 (2013).

20. Spychalski, W. L., Pisarek, M. & Szoszkiewicz, R. Microscale Insight into Oxidation of Single MoS$_2$ Crystals in Air. *J. Phys. Chem. C* **121**, 26027–26033 (2017).

21. Walter, T. N., Kwok, F., Simchi, H., Aldosari, H. M. & Mohney, S. E. Oxidation and oxidative vapor-phase etching of few-layer MoS$_2$. *J. Vac. Sci. Technol. B* **35**, 021203 (2017).

22. Kumar, P. *et al.* Direct visualization of out-of-equilibrium structural transformations in atomically thin chalcogenides. *npj 2D Mater. Appl.* **4**, 1–10 (2020).

23. Liu, H., Han, N. & Zhao, J. Atomistic insight into the oxidation of monolayer transition metal dichalcogenides: from structures to electronic properties. *RSC Adv.* **5**, 17572–17581 (2015).

24. Voronine, D. V. *et al.* Probing nano-heterogeneity and aging effects in lateral 2D heterostructures using tip-enhanced photoluminescence. *Opt. Mater. Express* **9**, 1620–1631 (2019).

25. Gao, J. *et al.* Aging of Transition Metal Dichalcogenide Monolayers. *ACS Nano* **10**, 2628–2635 (2016).

26. Tongay, S. *et al.* Broad-range modulation of light emission in two-dimensional semiconductors by molecular physisorption gating. *Nano Lett.* **13**, 2831–2836 (2013).

27. Liang, T., Sawyer, W. G., Perry, S. S., Sinnott, S. B. & Phillpot, S. R. Energetics of oxidation in $MoS_2$ nanoparticles by density functional theory. *J. Phys. Chem. C* **115**, 10606–10616 (2011).

28. Park, S. *et al.* Effect of Adventitious Carbon on Pit Formation of Monolayer $MoS_2$. *Adv. Mater.* **32**, 2003020 (2020).

29. Ko, T. Y. *et al.* On-stack two-dimensional conversion of $MoS_2$ into $MoO_3$. *2D Mater.* **4**, 014003 (2016).

30. Zhu, H. *et al.* Remote Plasma Oxidation and Atomic Layer Etching of $MoS_2$. *ACS Appl. Mater. Interfaces* **8**, 19119–19126 (2016).

31. Liu, Y. *et al.* Layer-by-layer thinning of $MoS_2$ by plasma. *ACS Nano* **7**, 4202–4209 (2013).

32. Alam, M. H. *et al.* Wafer-Scalable Single-Layer Amorphous Molybdenum Trioxide. *ACS Nano* **16**, 3756–3767 (2022).

33. Kang, S. *et al.* Enhanced Photoluminescence of Multiple Two-Dimensional van der Waals Heterostructures Fabricated by Layer-by-Layer Oxidation of $MoS_2$. *ACS Appl. Mater. Interfaces* **13**, 1245–1252 (2021).

34. Chu, X. S., Li, D. O., Green, A. A. & Wang, Q. H. Formation of $MoO_3$ and $WO_3$ nanoscrolls from $MoS_2$ and $WS_2$ with atmospheric air plasma. *J. Mater. Chem. C* **5**, 11301–11309 (2017).

35. Rao, R., Islam, A. E., Campbell, P. M., Vogel, E. M. & Maruyama, B. In situ thermal oxidation kinetics in few layer $MoS_2$. *2D Mater.* **4**, 025058 (2017).

36. Ross, F. M. & Gibson, J. M. Dynamic Observations of Interface Propagation during Silicon Oxidation. *Phys. Rev. Lett* **68**, 1782–1786 (1992).

37. Chithambararaj, A. & Bose, A. C. Investigation on structural, thermal, optical and sensing properties of meta-stable hexagonal $MoO_3$ nanocrystals of one dimensional structure. *Beilstein J. Nanotechnol* **2**, 585–592 (2011).

38. Gatan Inc. Molybdenum | EELS Atlas. *2022* https://eels.info/atlas/molybdenum.

39. Lajaunie, L., Boucher, F., Dessapt, R. & Moreau, P. Quantitative use of electron energy-loss spectroscopy Mo-$M_{2,3}$ edges for the study of molybdenum oxides. *Ultramicroscopy* **149**, 1–8 (2015).

40. Li, Y. *et al.* Oxygen vacancy-rich $MoO_{3-x}$ nanobelts for photocatalytic $N_2$ reduction to $NH_3$ in pure water. *Catal. Sci. Technol.* **9**, 803–810 (2019).

41. Wang, D., Su, D. S. & Schlögl, R. Electron Beam Induced Transformation of $MoO_3$ to $MoO_2$ and a New Phase MoO. *Zeitschrift für Anorg. und Allg. Chemie* **630**, 1007–1014 (2004).

42. Kim, J. H. *et al.* van der Waals epitaxial growth of single crystal α-$MoO_3$ layers on layered materials growth templates. *2D Mater.* **6**, 015016 (2018).

43. Molina-Mendoza, A. J. *et al.* Centimeter-Scale Synthesis of Ultrathin Layered $MoO_3$ by van der Waals Epitaxy. *Chem. Mater.* **28**, 4042–4051 (2016).

44. Yoon, A., Kim, J. H., Yoon, J., Lee, Y. & Lee, Z. Van der Waals Epitaxial Formation of Atomic Layered α-$MoO_3$ on $MoS_2$ by Oxidation. *ACS Appl. Mater. Interfaces* **12**, 22029–22036 (2020).

45. Yan, B. *et al.* Orientation controllable growth of $MoO_3$ nanoflakes: Micro-raman, field emission, and birefringence properties. *J. Phys. Chem. C* **113**, 20259–20263 (2009).

46. Zeng, H. C., Sheu, C. W. & Hia, H. C. Kinetic Study of Vapor-Phase Preparation of Orthorhombic Molybdenum Trioxide. *Chem. Mater.* **10**, 974–979 (1998).

47. Lopatin, S. *et al.* Optimization of monochromated TEM for ultimate resolution imaging and ultrahigh resolution electron energy loss spectroscopy. *Ultramicroscopy* **184**, 109–115 (2018).

48. Park, J. *et al.* Bandgap measurement of thin dielectric films using monochromated STEM-EELS. *Ultramicroscopy* **109**, 1183–1188 (2009).

49. Fjellvåg, Ø. S. *et al.* Crystallization, Phase Stability, and Electrochemical Performance of β-$MoO_3$ Thin Films. *Cryst. Growth Des.* **20**, 3861–3866 (2020).

50. Hanson, E. D. *et al.* Systematic Study of Oxygen Vacancy Tunable Transport Properties of Few-Layer $MoO_{3-x}$ Enabled by Vapor-Based Synthesis. *Adv. Funct. Mater.* **27**, 1605380 (2017).

51. Peña, F. de la *et al.* hyperspy/hyperspy: Release v1.7.3. (2022) doi:10.5281/ZENODO.7263263.

52. Reidy, K. *et al.* Direct imaging and electronic structure modulation of moiré superlattices at the 2D/3D interface. *Nat. Commun.* **12**, 1290 (2021).

53. Schneider, G. F., Calado, V. E., Zandbergen, H., Vandersypen, L. M. K. & Dekker, C. Wedging transfer of nanostructures. *Nano Lett.* **10**, 1912–1916 (2010).

# Supplemental Information: Kinetic control for planar oxidation of MoS$_2$

Kate Reidy[1], Wouter Mortelmans[1], Seong Soon Jo[1], Aubrey Penn[2], Baoming Wang[1], Alexandre Foucher[1], Frances M. Ross[1], R. Jaramillo[1]

1. Department of Materials Science and Engineering, Massachusetts Institute of Technology, Cambridge, MA, USA

2. MIT.nano, Massachusetts Institute of Technology, Cambridge, MA, USA

## 1. Thermal oxidation at lower temperature and texture of crystalline α-MoO$_3$

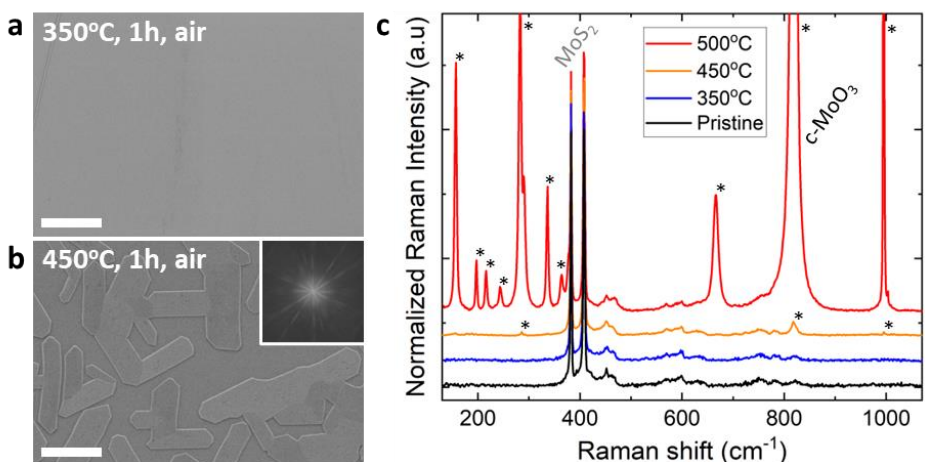

**Figure S1:** SEM and Raman of thermal oxidation of MoS$_2$ at lower temperatures. **(a-b)** SEM image and Fourier transformation (inset) of thermal oxidation of bulk MoS$_2$ at respectively 350 and 450 °C for 1 hr in ambient air. At 350 °C, no formation of MoO$_3$ is visible, while at 450 °C we observe textured, crystalline MoO$_3$ crystals. **(c)** Raman of thermal oxidation of MoS$_2$ showing the onset of apparent thermal oxidation at 450 °C.

## 2. Comparison of EELS Mo-M and O-K edge data with database spectra for $MoO_3$

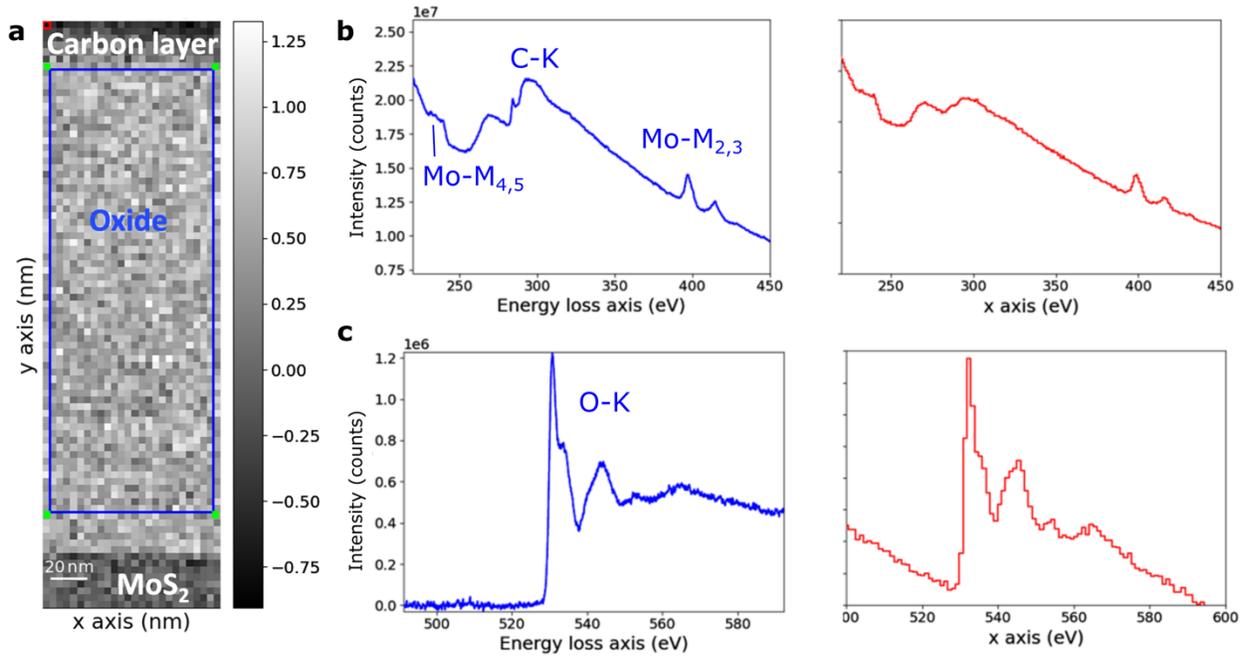

**Figure S2:** Experimental EELS data compared to reference spectra for $MoO_3$ particles on holey C film, from EELS Atlas.[1] **(a)** EELS intensity map where the blue box represents the area that the EELS spectrum in **Fig. 1g** is averaged over, **(b)** Molybdenum M edges and amorphous carbon K-edge (originating from residual carbon contamination and the amorphous FIB carbon protection layer) compared to reference $MoO_3$ spectra. **(c)** Oxygen K edge. The close spectral match for both edges confirms the bonding and oxidation states of Mo and O in $MoO_3$. Other oxides, such as $MoO_2$, exhibits notable differences in the intensity and position of these peaks.[2]

## 3. XPS data showing lack of oxidation of MoS$_2$ at 350 °C

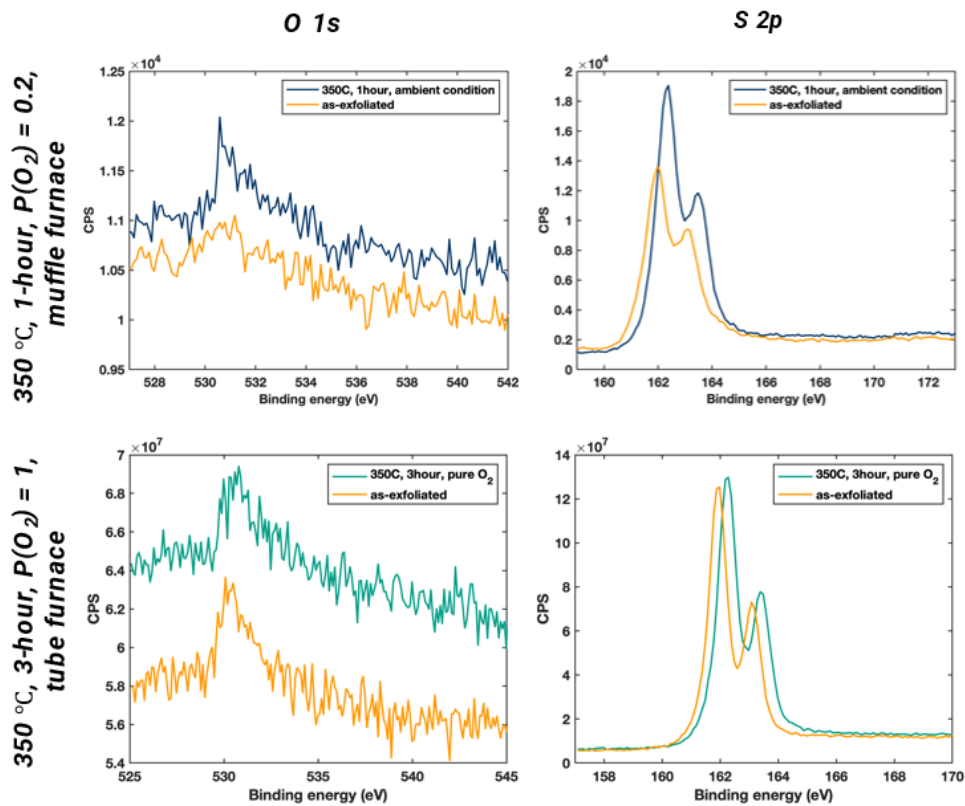

**Figure S3:** XPS data measured on MoS$_2$ after 1 hr annealing in a muffle furnace in lab ambient air (top), and in a tube furance with 1 atm O2 gas (bottom). The O 1s signals are not appreciably changed by either annealing treatment, suggesting that no oxide layer has formed. The S 2p signals are affected by annealing; presumably due to a small increase in sulfur vacancy concentration on annealing.

## 4. Thickness variation and roughness of plasma-processed amorphous MoO$_3$

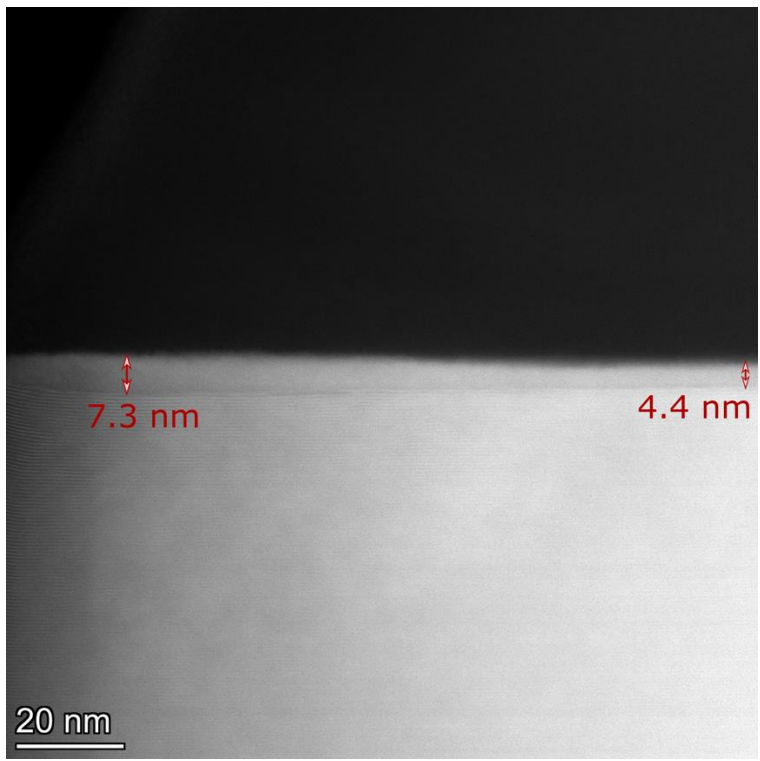

**Figure S4:** Cross sectional HAADF STEM image of MoS$_2$ after 30 min plasma oxidation. Arrows indicate the thickness of the oxide, which varies by a few nm across the sample area, in accordance with AFM roughness measurements.

## 5. Spectroscopic ellipsometry: Raw data and model refinement

The raw spectroscopic ellipsometry data of the plasma oxidation process (250 W and 3 sccm) with varying time is presented in **Fig S5a**. The data is measured at incident angles of 65 and 70° but only the angle of 70° is presented for clarity. Psi (Ψ) and delta (Δ) are displayed at the top, where it is observed that delta is very sensitive to the oxidation time (and thickness). The pseudo-$n$ and $k$ values and depolarization are presented in respectively the middle and bottom panels. The depolarization is below 5% for all measurements. In **Fig S5b**, the fitting results for the samples in (a) are presented including the oxide thickness and mean squared error (MSE) (top), the A$_n$ and B$_n$ dielectric constants (middle) and A$_k$ and B$_k$ extinction coefficients of the Cauchy model (bottom). The MSE for all samples is below 2 indicating a good fit. The uncertainty on the optical parameters of the oxide is rather large, hence no conclusions are made on the impact of oxidation processing conditions on the oxide optical properties.

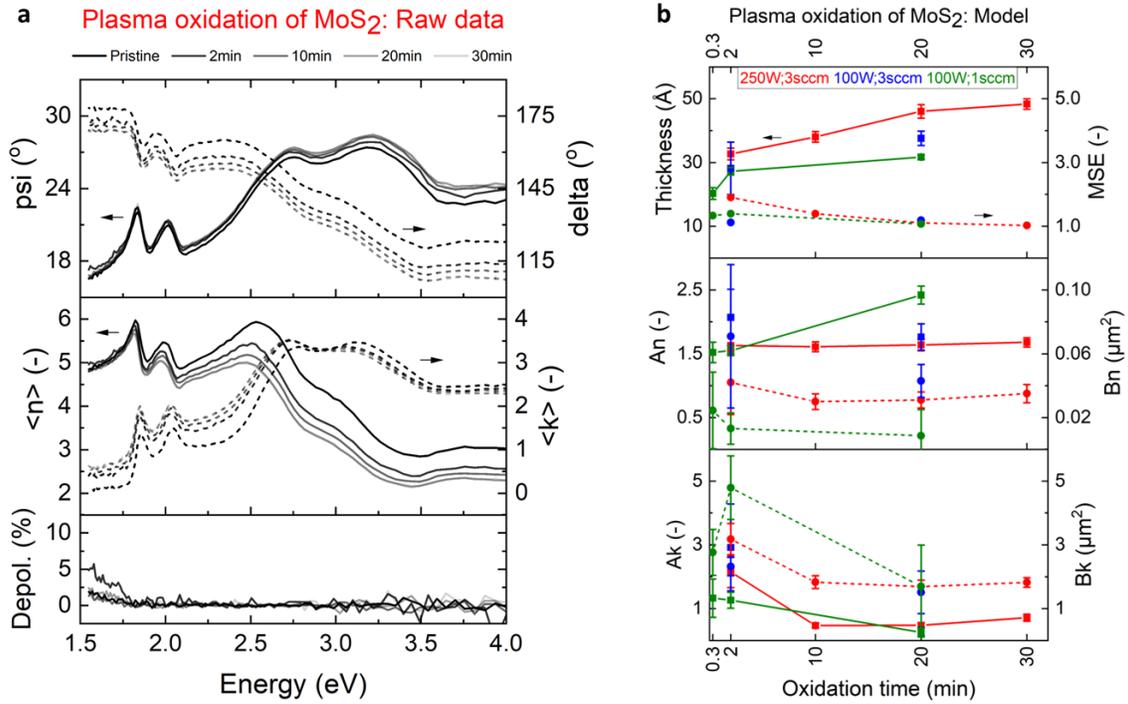

**Figure S5**: Spectroscopic ellipsometry data and model parameters of $MoS_2$ plasma oxidation. **(a)** SE data showing psi and delta (top), pseudo *n* and *k* (middle) and depolarization (bottom) for $MoS_2$ plasma-oxidized at 250 W and 3 sccm $O_2$ for various durations. **(b)** Results of SE modelling and fitting, showing the Cauchy oxide thickness and mean-squared-error (top), Cauchy dielectric constant parameters $A_n$ and $B_n$ (middle), and Cauchy extinction coefficients $A_k$ and $B_k$. Square symbols and solid lines are for the left axes, circular symbols and dashed lines are for the right axes.

## References

1. Gatan Inc. Molybdenum | EELS Atlas. *2022* https://eels.info/atlas/molybdenum.

2. Lajaunie, L., Boucher, F., Dessapt, R. & Moreau, P. Quantitative use of electron energy-loss spectroscopy Mo-$M_{2,3}$ edges for the study of molybdenum oxides. *Ultramicroscopy* **149**, 1–8 (2015).